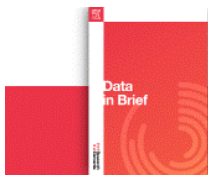

# ARTICLE INFORMATION

**Article title**

A Multimodal Tiered Magnetic Polarity Inversion Features Dataset for Space Weather Forecasting

**Authors**

Ziba Khani *, Anli Ji, Manolis K. Georgoulis, Berkay Aydin

**Affiliations**

Department of Computer Science, Georgia State University, Atlanta, Georgia, USA

Applied Physics Lab, Johns Hopkins University

**Corresponding author's email address and Twitter handle**

zkhani1@student.gsu.edu

**Abstract**

This article presents a publicly available, multimodal, tiered magnetic polarity inversion lines (MPILs) dataset extracted from the Solar Dynamics Observatory's (SDO) Helioseismic and Magnetic Imager (HMI) Active Region Patches (SDO/HMI HARP) between May 2010 and April 2025. The dataset comprises four distinct tiers, each generated by running our detection methodology with four different magnetic field strength thresholds to capture nuanced variations in MPIL features at multiple levels of detail. In total, we provide 6,695 HARP series mapped using the Lambert Cylindrical Equal Area (CEA) projection at a 12-minute cadence. This tiered approach ensures that each tier captures specific sensitivities to polarity changes, enabling researchers to tailor their analyses to a range of scientific and operational objectives. In each threshold tier, we offer six binary MPIL masks associated with heliophysics and space weather forecasting, including MPIL, Region of Polarity Inversion (RoPI), positive/negative polarity regions, unsigned polarity regions, and MPIL convex hulls. Furthermore, structured metadata in the form of multivariate time series is included, allowing users to track and analyze MPIL properties over time.

# SPECIFICATIONS TABLE

| Subject | Solar Physics, Computer Science Applications |
|---|---|
| Specific subject area | Solar magnetic polarity inversion lines, Space weather forecasting |
| Type of data | HDF5 (containing binary masks and numerical metadata), Multimodal data, Timeseries, Binary rasters |
| Data collection | The dataset is generated from Helioseismic and Magnetic Imager (HMI) onboard NASA's Solar Dynamics Observatory (SDO), line-of-sight magnetogram patches at a 12-minute cadence. The data generation process involved automated detection of magnetic polarity inversion lines (MPILs) using thresholding and morphological operations. Quality control included noise filtering, size-based exclusion criteria, and normalization to ensure consistency. The dataset integrates spatio-temporal characteristics of polarity inversion lines, which can be beneficial for forecasting events such as solar flares. |
| Data source location | Data collected from Helioseismic and Magnetic Imager (HMI) onboard NASA's Solar Dynamics Observatory (SDO). The dataset is stored at Harvard Dataverse and maintained by DMLAB at Georgia State University. |
| Data accessibility | **Repository name:** Harvard Dataverse<br><br>**Data identification number:** DOI: [10.7910/DVN/BKP1RH]<br><br>**Direct URL to data:** [https://doi.org/10.7910/DVN/BKP1RH]<br><br>**Instructions for accessing these data:** The dataset is open access and can be downloaded from the DOI link. Each HDF5 file contains MPIL rasters and extracted metadata. |
| Related research article | Ji, A.; Cai, X.; Khasayeva, N.; Georgoulis, M. K.; Martens, P. C.; Angryk, R. A. & Aydin, B., "A Systematic Magnetic Polarity Inversion Line Data Set from SDO/HMI Magnetograms". *The Astrophysical Journal Supplement Series, The American Astronomical Society,* **2023**, *265*, 28 available at [DOI 10.3847/1538-4365/acb43a] |

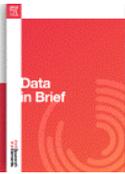

## VALUE OF THE DATA

This comprehensive dataset is available at [DOI: 10.7910/DVN/BKP1RH], serving as a valuable resource for space weather forecasting, solar physics research, and machine learning applications. Researchers can leverage their multiple tiers and associated metadata to gain deeper insights into the complex dynamics of active regions and refine solar event prediction strategies. The dataset supports exploration of solar magnetic field evolution and active region behavior, with applications across both observational and predictive domains. Below, we highlight the key attributes that make this dataset especially useful to the scientific community.

- **Comprehensive Solar Cycle Coverage:**
  This dataset spans Solar Cycle 24 and partially the ongoing Cycle 25 (2010–2025), providing a long-term, high-cadence (12-minute) record of Magnetic Polarity Inversion Lines (MPILs) derived from SDO/HMI Active Region Patches (HARP). The availability of MPIL masks and metadata enables researchers to study long-term solar magnetic field evolution and its relationship with space weather events.

- **Multimodal Tiered MPIL Masks and Features:** Each HDF5 file contains six binary masks representing MPILs, Regions of Polarity Inversion (RoPI), positive/negative polarity, unsigned polarity, and convex hulls. Additionally, metadata includes quantitative shape descriptors and MPIL size and count, making it a valuable dataset for studying solar active region complexity.

- **Facilitates Machine Learning-Based Solar Transient Event Prediction:** The dataset can be integrated with historical transient event metadata (such as flares or coronal mass ejections), allowing researchers to apply supervised learning models for space weather forecasting. The structured, spatiotemporal nature of the dataset supports spatial/temporal deep learning models, such as recurrent or transformer-based networks, for time series analysis.

- **Standardized Data Format for Reusability:** The dataset is stored in HDF5 format, ensuring efficient data handling, compression, and accessibility across various programming environments. Researchers can easily extract MPIL masks and associated metadata for their analyses, supporting reproducibility and interoperability with existing space weather models.

- **Applicable to Multiple Research Areas:** Beyond space weather forecasting, the dataset can be used in studies of solar magnetic field morphology, polarity evolution, active region complexity analysis, and other event precursors.

- **Supports Operational Space Weather Forecasting:** This dataset provides essential MPIL-related features that can be integrated into real-time space weather monitoring systems. By leveraging the sliding window approach, researchers can simulate continuous data assimilation techniques for transient event prediction models used in space weather forecasting operations.

# BACKGROUND

The Sun's magnetic field is highly dynamic, forming active regions, sunspots, and flares. Magnetic Polarity Inversion Lines (MPILs) are critical derived features that separate regions of opposite magnetic polarity. These shear layers play a fundamental role in space weather forecasting, as they have been identified as precursors to major solar events such as solar flares and coronal mass ejections (CMEs). The complexity of MPILs within an active region is strongly correlated with energy buildup, making them a crucial feature for space weather analysis [1].

Various MPIL detection methods exist, using intensity thresholds, segmentation, or machine learning-based classification [2][3][4]. However, most prior works present their detection methodology but lack data accessibility, limiting large-scale studies and operational forecasting applications.

To address this gap, we introduce a publicly available MPIL dataset derived from SDO/HMI Active Region Patches (HARP) spanning May 2010–April 2025 [5]. This dataset provides structured, multimodal MPIL data in HDF5 format, facilitating space weather modeling, solar magnetic field studies, and deep learning applications.

# DATA DESCRIPTION

This section describes the dataset structure, including all folders, subfolders, and files, to ensure clarity for the reader.

**Dataset Overview**

The dataset consists of 4,325,011 individual active region patches derived from the Solar Dynamics Observatory/Helioseismic and Magnetic Imager (SDO/HMI) Line-of-Sight (LoS) magnetograms. The data spans May 2010 to April 2025, covering the entire Solar Cycle 24 and partially covering Solar Cycle 25, to date.

Each dataset instance is stored in HDF5 (Hierarchical Data Format 5) files, structured as follows:

- **Binary Masks (six types)**:

    1. **MPIL Mask:** Primary mask for polarity inversion lines.
    2. **Regions of Polarity Inversion (RoPI):** Areas where opposite polarity regions meet.
    3. **Positive Polarity Region:** Areas with positive magnetic field strength.
    4. **Negative Polarity Region:** Areas with negative magnetic field strength.
    5. **Unsigned Polarity Region:** Combined representation of positive and negative regions.
    6. **Convex Hull of MPILs:** Enclosing a convex hull around the MPILs.

- **Multivariate Time Series Features**:

    1. **MPIL Count:** Total number of disconnected MPILs.

2. **MPIL Size:** Number of pixels in the MPIL mask (px).
3. **Total Perimeter of MPIL Contours:** The perimeter of the area enclosed by the MPIL (px)
4. **Masked Unsigned Flux of RoPI:** Sum of $|B_{los}|$ enclosed by RoPIs (Gauss)
5. **Masked Unsigned Flux of MPIL:** Sum of $|B_{los}|$ enclosed by MPIL (Gauss)
6. **RoPI Area:** Number of pixels in the regions of polarity inversion (px).
7. **RoPI Total Contour Area:** Total area enclosed by the regions of polarity inversion (px)
8. **RoPI Total Contour Perimeter:** Total perimeter of RoPI contours (px)
9. **Convexity Ratio:** This shape descriptor represents the ratio between the perimeter of the MPIL area's convex hull and the total sum of pixels in the MPILs. A higher convexity ratio suggests a more convex shape.
10. **Eigenvalues of the MPIL covariance matrix:** Describe the distribution of pixel positions along the principal axes of the MPIL. The first (larger) eigenvalue reflects variance along the major axis, and the second (smaller) along the minor axis, indicating the feature's elongation.
11. **Hu Moments:** A set of seven invariant shape descriptors derived from image moments, capturing properties such as symmetry, elongation, and complexity. These are invariant to image translation, rotation, and scaling, making them suitable for robust shape characterization of solar features [6].
12. **Minimum aspect ratio of PILs**: The smallest aspect ratio between width and height of MPILs, indicating their elongation.
13. **Ratio of RoPI to the Minimum Orthogonal Bounding Rectangle (MOBR):** The fraction of RoPI area relative to its smallest upright (axis-aligned) bounding rectangle.
14. **Ratio of RoPI to the Minimum Rotated Bounding Rectangle (MRBR):** The fraction of RoPI area relative to its smallest rotated bounding rectangle.
15. **Ratio of RoPI to the Minimum Bounding Ellipse (BE):** The fraction of RoPI area relative to its best-fitting minimum bounding ellipse.
16. **Ratio of RoPI to its Convex Hull area (CH):** The fraction of RoPI area relative to its convex hull, measuring shape compactness.

## EXPERIMENTAL DESIGN, MATERIALS, AND METHODS

In this section, we will summarize the steps involved in generating MPILs. We will also provide visualizations for a running example using the MPIL detection results for HMI Active Region Patch 3772 on 2014 February 17 at 15:00:00 UT. The visualizations of this active region cutout are shown in Fig. 2.

### PIL Detection

**Detecting the Polarity Regions:** A flowchart in **Fig. 1** illustrates our extended noise-aware process for Magnetic Polarity Inversion Line (MPIL) detection, which consists of multiple steps, beginning with creating binary masks for positive and negative polarity regions. These masks are generated by applying a threshold for magnetic field strength, followed by a preliminary filter to eliminate noise-like smaller regions.

To identify the regions of polarity inversion (RoPI), we spatially buffer the positive and negative polarity regions using morphological dilation. The potential RoPI is then defined as the intersection between these buffered regions where positive and negative polarity areas are adjacent.

At this stage, a field strength magnitude filter is applied to potential RoPIs to filter out those with weaker magnetic field strength values around the separating regions. Morphological thinning is then performed on the remaining RoPIs to produce potential MPILs. Finally, the most relevant MPILs are established by enforcing an MPIL size threshold and filling the gaps between the closest MPILs. Subsequent sections will provide a detailed explanation of each of these stages.

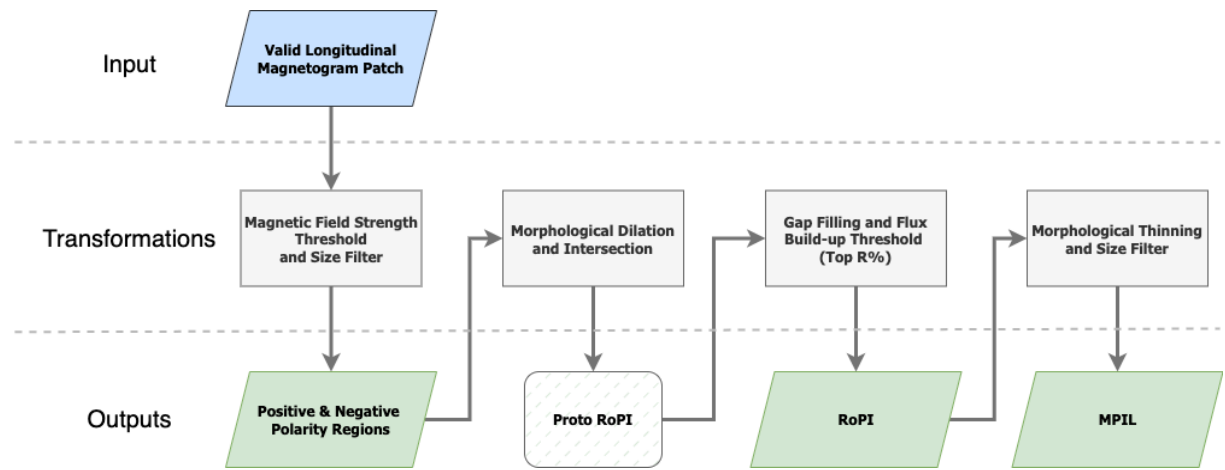

**Figure 1:** Overall workflow of our MPIL detection method.

To locate areas of positive and negative polarity, it is essential to consider the uncertainties that may arise from the instruments and models used to generate magnetic field strength maps. Therefore, for each tier of the data, we initially applied a strength filter to the original maps with the same absolute value of ±50, ±100, ±250, and ±500 G for both polarities but different signs to identify the regions with the most robust magnetic fields. The threshold for the unsigned field strength is denoted as $S_{th}$, and we determine the positive and negative polarity regions based on the magnetic field strength (in this case, line-of-sight magnetic field $B_{los}$) using the following expressions:

$$Positive = \begin{cases} 1 & \text{if } B_{los} \geq S_{th} \\ 0 & \text{if } otherwise \end{cases} \quad (1)$$

$$Negative = \begin{cases} 1 & \text{if } B_{los} \leq -S_{th} \\ 0 & \text{if } otherwise \end{cases} \quad (2)$$

The map of positive polarity regions is represented as a binary raster where the value of 1 is assigned to areas with a magnetic field strength greater than or equal to $S_{th}$, and 0 is assigned to the rest of the region.

Similarly, the map of negative polarity regions is constructed by assigning the value of 1 to areas with magnetic field strength less than or equal to $-S_{th}$, and the value of 0 to the remaining regions. Additionally, we remove small connected components in this step to eliminate potentially noisy structures in polarity regions. This removal routine employs a polarity region minimum connected component size parameter of 14 pixels (i.e.,~1.86 $Mm^2$ of the photospheric area) and filters small groups. These polarity region binary maps are utilized in the subsequent steps to identify the boundaries of the regions with strong polarity. We demonstrate the positive and negative polarity region maps in **Fig. 2 (a-c)**.

**Identifying the Regions of Polarity Inversion (RoPI):** The RoPI is identified by the regions where the positive and negative polarity regions are adjacent. To identify these areas, we spatially expand the positive and negative polarity regions using morphological dilation [7], a basic operation in mathematical morphology used to expand the boundaries of shapes in an image. It involves sliding a structuring element (i.e., a kernel) over each pixel in the image and replacing each pixel with the maximum value within the area covered by the structuring element. The structuring element's size and shape determine the extent of the dilation. Although the kernel size can vary, we have utilized a 4 by 4 binary structuring element (pixels in CEA) for our PIL datasets. Then, we generate a RoPI binary map by checking whether the binary values of the dilated region in the positive polarity region intersect with the negative polarity region, as shown in **Eq. 3**:

$$ProtoRoPI = \begin{cases} 1 & \text{if } Pde_i \cap Nde_j \neq \emptyset \\ 0 & \text{otherwise} \end{cases} \qquad (3)$$

where **Pde** and **Nde** are the pixels of dilated positive and negative regions.

**RoPI Field Strength Magnitude Filter:** The overlapping areas of expanded polarity regions generate several fragmented regions. These separate parts within the polarity inversion regions have associated magnetic field strengths. To account for potential discontinuities in RoPI segmentation, we first apply a gap-filling operation to bridge small separations between adjacent RoPI components that likely belong to the same structure but were initially disconnected due to noise or minor variations in the data. This ensures a more continuous representation of RoPIs before computing their cumulative magnetic field strength. We consider these areas distinct entities and compute the total unsigned magnetic flux, or the absolute sum of magnetic field strengths, for each continuous RoPI element (i.e., we use the connected components from the RoPI raster and calculate the total unsigned flux). Next, we organize the RoPI components in descending order based on their magnetic field strength percentage (concerning the total unsigned flux covered by all RoPI). We then maintain the RoPIs whose cumulative sum is below a predetermined threshold of *R%*. This filtering process ensures that the selected polarity inversion lines (PILs) represent at least *R%* of the magnetic flux among the pixels found in the RoPI, thereby eliminating relatively minor flux MPILs.

$$RoPI = \begin{cases} 1 & \text{if } \text{cumsum}\left(\dfrac{|B_{los_i}|}{\sum_{i=1}^{q} |B_{los}|}\right) \leq R\% \\ 0 & \text{otherwise.} \end{cases} \qquad (4)$$

In our example, we present the detected RoPI raster in Fig. 2 (d), where the RoPIs represent over 95% of the accumulated flux.

**Morphological Thinning and MPIL Size Filter:** We apply a morphological thinning operation on the RoPIs generated and filtered in the previous steps. Morphological thinning [8], commonly employed in image processing for skeletonization, removes most original foreground pixels while maintaining the topology of the initial shapes. Each RoPI is then converted into pixel-chain representations of candidate MPILs. Next, we refine these MPILs by implementing a size threshold. MPIL size is determined by the number of contiguous pixels in each disconnected MPIL. Ultimately, we retain MPILs with a size equal to or greater than ($L_{th}$) and discard those smaller than ($L_{th}$). In **Fig. 2 (e)**, we demonstrate the MPILs detected after this step.

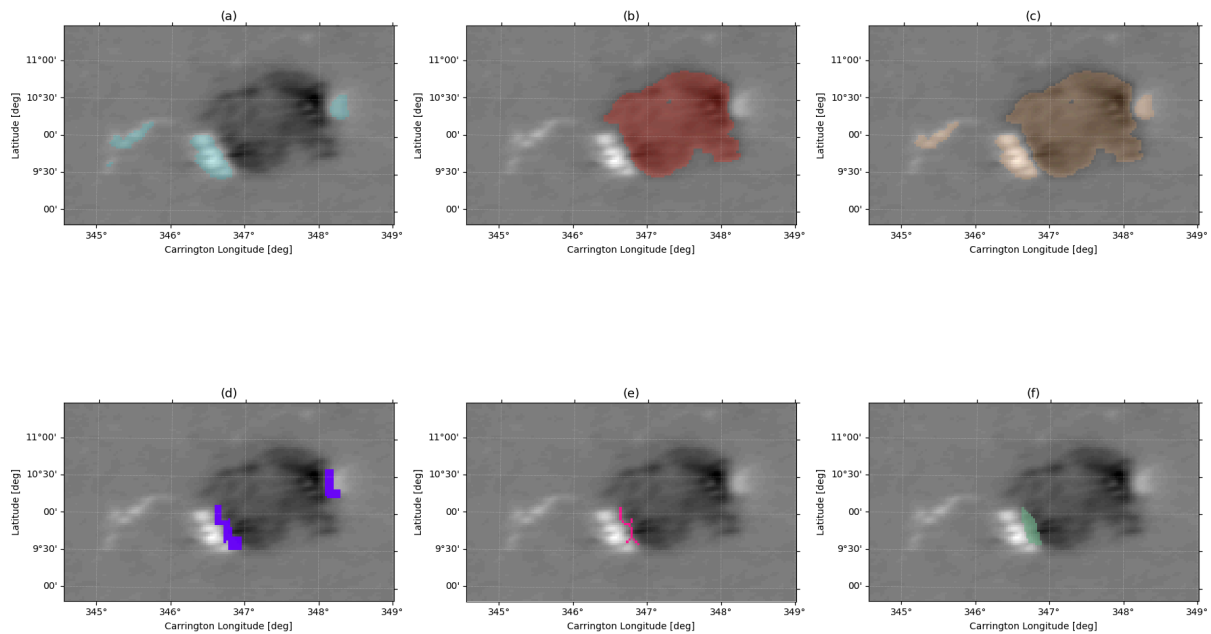

**Fig 2:** Binary masks overlaying the HMI magnetogram patch of HMI Active Region Patch 3772 on 2014 February 17 at 15:00:00 UT. In (a), the blue area shows the positive polarity region. In (b), the red area shows the negative polarity region. In(c), the orange area represents the masked unsigned polarity region, i.e., the union of positive and negative polarities. In (d), the purple area indicates the RoPI binary mask. In (e), the pink line represents the detected MPIL binary mask. In (f), the green area shows the binary mask of the convex hull of the MPIL network in the active region.

# Computational Framework

- **Software**: C++ (GPU-accelerated processing)

- **Data Sources**: SDO/HMI Active Region Patches (Line-of-Sight magnetogram components), GOES solar flare catalog for additional exploratory analysis.

## Exploratory Analysis

To enable flare-related analysis, we labeled each dataset instance by linking it with NOAA/GOES flare records. First, we mapped each HARP number in our dataset to its associated NOAA active region(s) using the HARP–active region associations. We then matched these active regions with flare event data, assigning to each timestamp the most intense flare occurring within the same 12-minute cadence interval. Instances without a matching flare were labeled as non-flaring (NF). We performed a set of exploratory analyses to characterize the relationships between magnetic polarity inversion line (MPIL) morphology and solar flare occurrence and intensity. Our focus is on three MPIL-derived features: MPIL size, MPIL count, and RoPI area, and their behavior with respect to flare strength. Flare events are grouped into six ordered categories based on their GOES X-ray classification: non-flaring (NF), C-class subdivided as C1–C4 and C5–C9, M-class as M1–M4 and M5–M9, and X1+ to retain both hierarchical ordering and intra-class granularity. In the main text, we present results for the baseline ±100 G threshold, which offers a balance between noise reduction and retention of mid-field structures. Equivalent plots and interpretations for the other thresholds (±50 G, ±250 G, and ±500 G) are provided in **Appendix A** for completeness.

**Analysis 1: Statistical distribution of MPIL features: Fig. 4** presents the distributions of MPIL size, MPIL count, and RoPI area stratified by flare class categories for the ±100 G threshold, using violin plots on a logarithmic scale. Related classes share hue families (C-class in greenish tones, M-class in orange) to emphasize progression, and the x-axis is ordered by increasing flare strength. The violins include internal density ticks to indicate sample concentration and spread; the log scaling reveals both the bulk of the data and the extended high-value tails associated with stronger events.

Qualitatively, all three MPIL-derived metrics exhibit a systematic upward shift with increasing flare class. Non-flaring instances are concentrated at the lower end of each feature's range (e.g., small MPIL sizes, few MPILs, small RoPI areas). In contrast, X-class flares preferentially populate the high-value tails. Intermediate classes (C and M subclasses) show gradual increases in central tendency and dispersion: for example, the transition from C1–C4 to C5–C9 is subtle, whereas M1–M4 and M5–M9 display more pronounced elevation, culminating in X1+ with the broadest and highest-valued distributions. Nonetheless, adjacent categories (especially upper C-class versus lower M-class) exhibit substantial overlap, indicating that no single feature provides perfect separation of flare classes. The violins also reveal that variance increases with flare strength: stronger flares arise from a broader range of MPIL configurations, suggesting greater heterogeneity in pre-flare MPIL morphology for high-intensity events.

These results establish that MPIL size, count, and the spatial extent of the polarity inversion region correlate positively with flare strength in a graded manner, but with overlapping distributions. These motivating downstream analyses incorporate temporal evolution and combinations of features for improved discrimination.

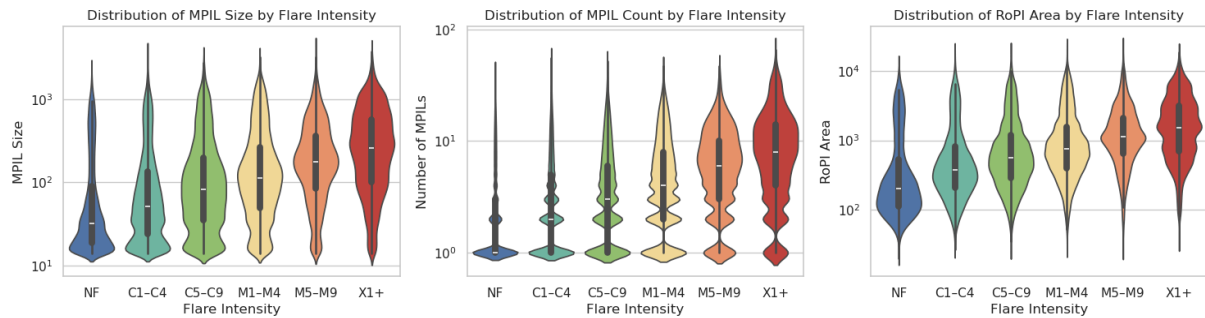

**Fig 4.** Violin plots of MPIL size (left), MPIL count (center), and RoPI area (right) across flare class categories for the ±100 G threshold. Distributions are shown on a logarithmic scale; white dots represent medians, thick bars indicate interquartile ranges, and internal ticks show local density. Related classes share hue families (C-class in green tones, M-class in orange) to emphasize progression. Stronger flares tend to have larger and more numerous MPILs and more extensive RoPI areas, though adjacent categories overlap substantially.

**Analysis 2: Histogram analysis between flare groups:** To quantify the separability of feature distributions across flare class categories, we computed pairwise Bhattacharyya distances [9] for the same three MPIL-derived features: MPIL size, MPIL count, and RoPI area, using the ±100 G threshold (**Fig. 5**). Distances were calculated separately for differences in mean and standard deviation. Larger values indicate greater divergence between two distributions.

The results show that separability is highest between the most distant categories (e.g., NF vs. X1+) and lowest between adjacent ones (e.g., C1–C4 vs. C5–C9, M1–M4 vs. M5–M9). MPIL Count consistently yields the largest distances, particularly between NF and high-intensity flares, indicating that the number of inversion lines is a strong differentiator of flare productivity. MPIL Size and RoPI area also demonstrate increasing distances with flare class difference, but the growth is more gradual. Standard deviation–based distances highlight that variability in these features also increases with flare strength, especially for the M- and X-class ranges. This suggests that higher-intensity flares are associated not only with higher median values of MPIL features but also with greater diversity in their morphological configurations.

Overall, the Bhattacharyya distance analysis reinforces the findings from the static distribution plots: MPIL features, especially MPIL Count, carry meaningful information for distinguishing between flare classes, though adjacent categories remain difficult to separate without incorporating additional temporal or multi-feature context.

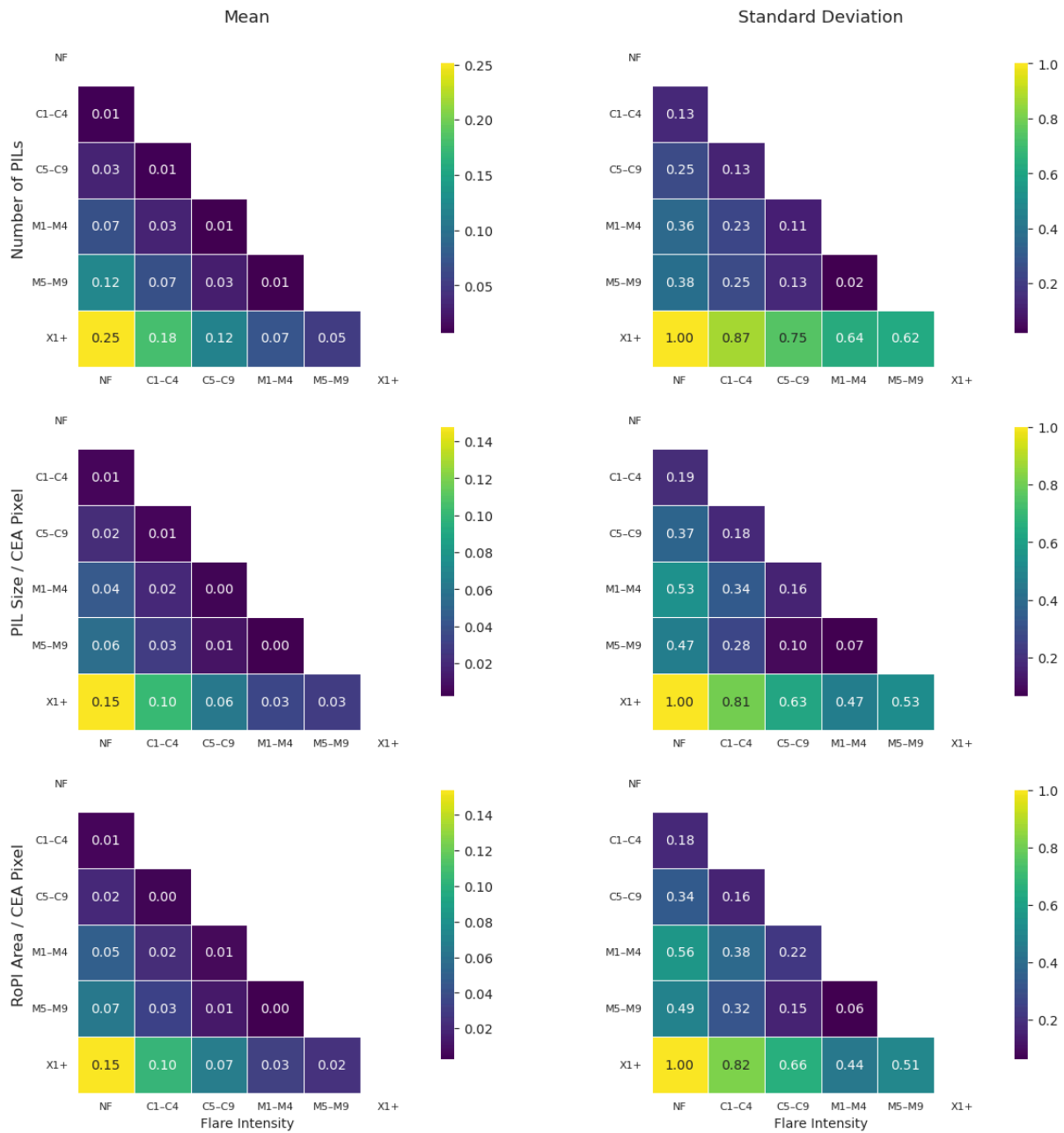

**Fig 5.** Pairwise Bhattacharyya distances between flare class categories for MPIL count (top row), MPIL size (middle row), and RoPI area (bottom row) in the ±100 G threshold tier. Left column: distances based on mean differences; right column: distances based on standard deviation differences. Higher values indicate greater distributional separation. MPIL Count provides the largest separations, particularly between NF and X-class flares, while standard deviation distances highlight increased feature variability at higher flare classes.

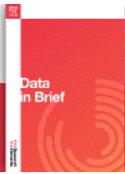

# LIMITATIONS

Although our data products highlight meaningful correlations between MPIL features and flare events, several important limitations should be acknowledged. First, our analysis is based on a fixed set of threshold values, and without the use of adaptive tuning or broader parameter sweeps, some PIL characteristics may be either underrepresented or overstated. Second, severe projection effects near the solar limbs can distort or even invalidate magnetograms, leading to partially or completely unused data in the PIL detection process. Finally, our uniform parameter settings—chosen for consistency—may not capture the full complexity of the underlying magnetic structures, leaving open the possibility that more refined or adaptive configurations could yield more accurate results.

# ETHICS STATEMENT

We confirm that we have read and follow the ethical requirements for publication in Data in Brief. We further confirm that our work does not involve human subjects, animal experiments, or data collected from social media platforms.

# CRediT AUTHOR STATEMENT

**Khani, Ziba:** Methodology, Software, Visualization, Writing Original draft preparation

**Ji, Anli**: Writing, Reviewing, and Editing

**Georgoulis, Manolis:** Supervision, Conceptualization

**Aydin, Berkay**: Supervision, Conceptualization, Validation, Writing, Reviewing, and Editing

# ACKNOWLEDGEMENTS

This work is supported under NSF Grant Award No.2104004. The authors would also like to thank the data teams from the NASA/SDO mission.

# DECLARATION OF COMPETING INTERESTS

The authors declare that they have no known competing financial interests or personal relationships that could have appeared to influence the work reported in this paper.

# Appendix A – Threshold Sensitivity of MPIL Features

In addition to the baseline ±100 G threshold presented in the main text, we evaluated MPIL feature distributions and separability for three alternative magnetic field strength thresholds: ±50 G, ±250 G, and ±500 G.
 The goal of this analysis is to examine how threshold selection affects the static morphology of MPILs and their relationship to flare class.
 For each threshold, we present:

1. **Static distribution plots** (violin plots) for MPIL Size, MPIL Count, and RoPI Area across flare class categories.

2. **Bhattacharyya distance matrices** quantifying pairwise separability between classes based on differences in means and standard deviations.

**Distribution of MPIL Features by Flare Class: Fig. A1** shows the distributions of MPIL size, MPIL count, and RoPI area for the ±50 G threshold.
 All three features display a clear upward trend in median and spread with flare strength.
 Compared to higher thresholds, feature magnitudes are generally larger, as the lower cutoff includes weaker-field PIL segments in addition to strong cores. Non-flaring cases are concentrated at the lower end of each feature's range, while X-class events dominate the high-value tails. Intermediate categories (C- and M-class subdivisions) show progressive increases in both central tendency and variability. The inclusion of weak-field structures increases the range of values within each class, slightly reducing class separation relative to ±100 G. Still, the overall positive correlation with flare class remains strong.

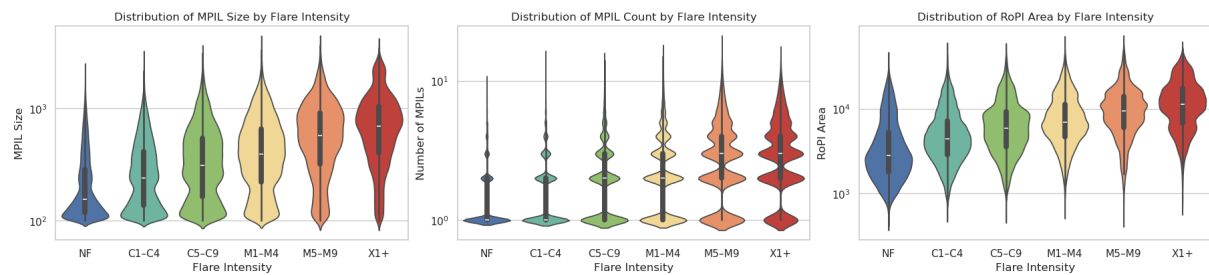

**Fig A1.** Violin plots of MPIL size (left), MPIL count (center), and RoPI area (right) across flare class categories for the ±50 G threshold. Distributions are shown on a logarithmic scale; white dots represent medians, thick bars indicate interquartile ranges, and internal ticks show local density. Lower thresholds capture both strong- and weak-field PIL structures, resulting in generally larger feature magnitudes.

**Bhattacharyya Distance Analysis: Fig. A2** presents Bhattacharyya distances for the ±50 G threshold.
 Separation is strongest between NF and X-class events for all three features, with MPIL Count yielding the highest distances overall. MPIL Size and RoPI Area also show meaningful separations, but with more gradual increases as the class gap widens.
 Standard deviation–based distances indicate that variability in all features increases substantially with flare strength, particularly for M- and X-class events.
These results suggest that while ±50 G captures more structural detail, the added weak-field content introduces noise that slightly reduces discrimination between adjacent flare classes.

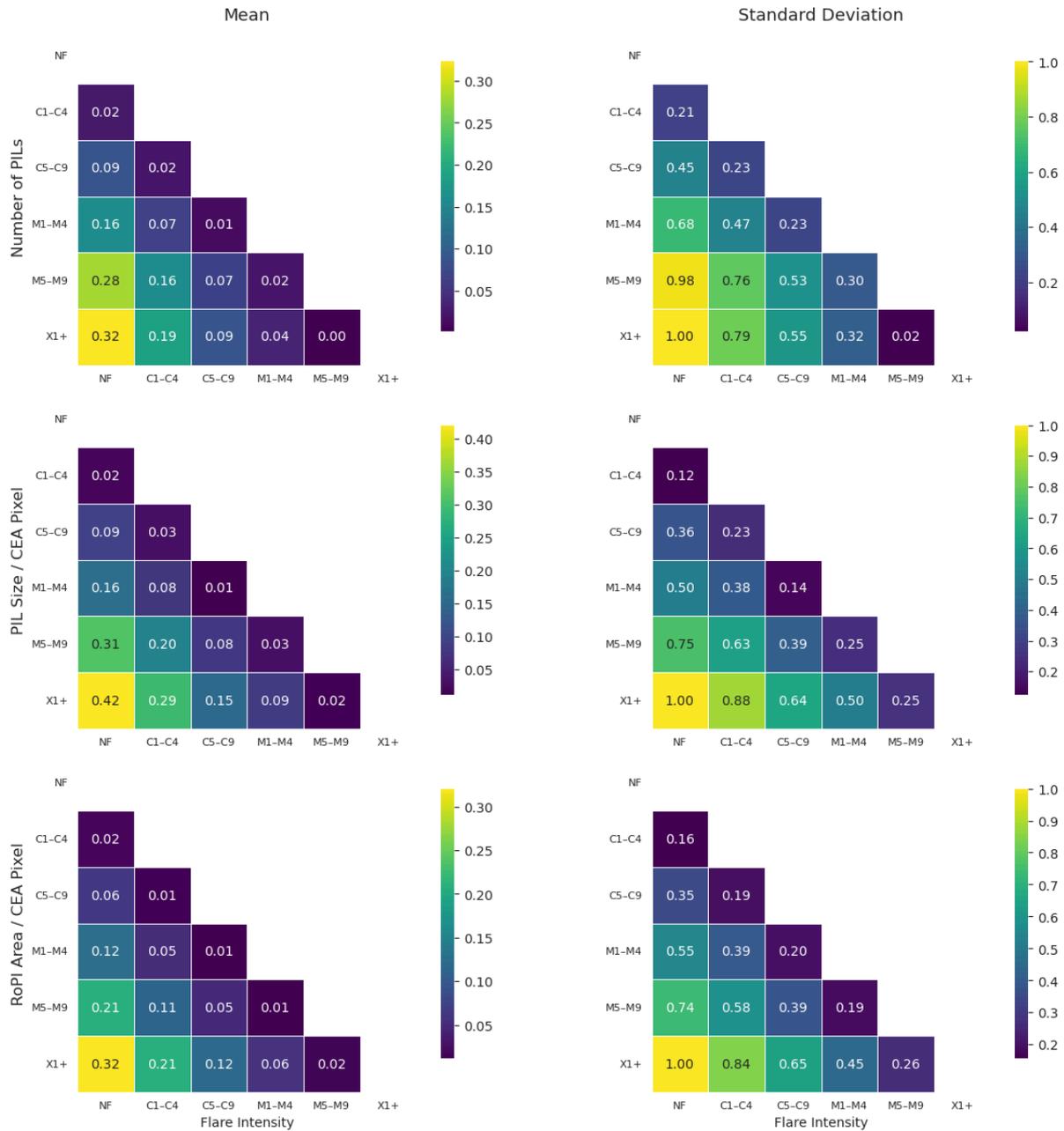

**Fig A2.** Pairwise Bhattacharyya distances between flare class categories for MPIL count (top row), MPIL size (middle row), and RoPI area (bottom row) in the ±50 G threshold tier. Left column: distances based on mean differences; right column: distances based on standard deviation differences. MPIL Count achieves the highest separations, particularly between NF and X-class flares, while standard deviation distances highlight increased variability at higher flare classes.

## A.2 ±250 G Threshold

**Distribution of MPIL Features by Flare Class: Fig. A3** shows the distributions of MPIL size, MPIL count, and RoPI area for the ±250 G threshold.

Compared to ±50 G and ±100 G, feature magnitudes are reduced because only the strongest-field MPIL cores are included. MPIL Size medians are compressed across all flare classes, with only a modest upward shift from NF to X1+. MPIL Count is low for all categories—most events have 1–2 PILs, resulting in minimal separation between flare classes. RoPI Area still shows a mild upward trend with flare class, but the differences are more minor than at lower thresholds.

Overall, the high threshold reduces variation and makes it harder to distinguish between flare classes based solely on static feature distributions.

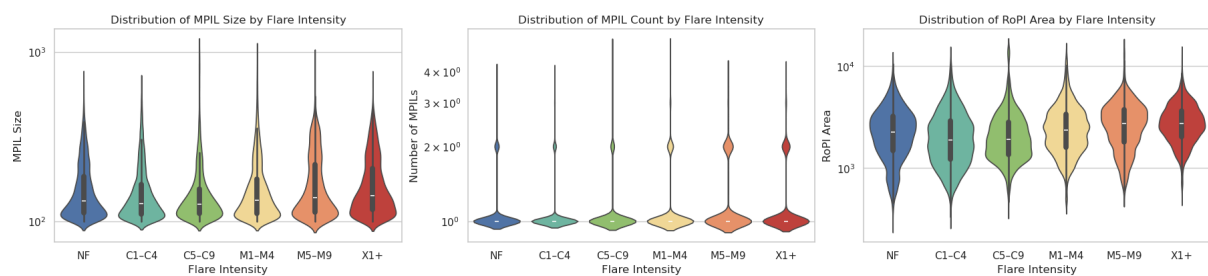

**Fig A3.** Violin plots of MPIL size (left), MPIL count (center), and RoPI area (right) across flare class categories for the ±250 G threshold. Distributions are shown on a logarithmic scale; white dots represent medians, thick bars indicate interquartile ranges, and internal ticks show local density. Higher thresholds reduce the range of feature values and diminish class separability.

**Bhattacharyya Distance Analysis: Fig. A4** presents Bhattacharyya distances for the ±250 G threshold.

Distances between classes are much smaller than for ±50 G and ±100 G, especially for MPIL Count, which shows little separability between adjacent or even distant categories.

RoPI Area provides the highest distances among the three features, though still far lower than at lower thresholds.

Standard deviation–based distances highlight that variability differences across classes are reduced, reflecting the fact that most weak- and moderate-field structures have been excluded by the higher cutoff. This suggests that while ±250 G isolates the most intense field regions, it also discards much of the flare-relevant variability found in mid-field structures.

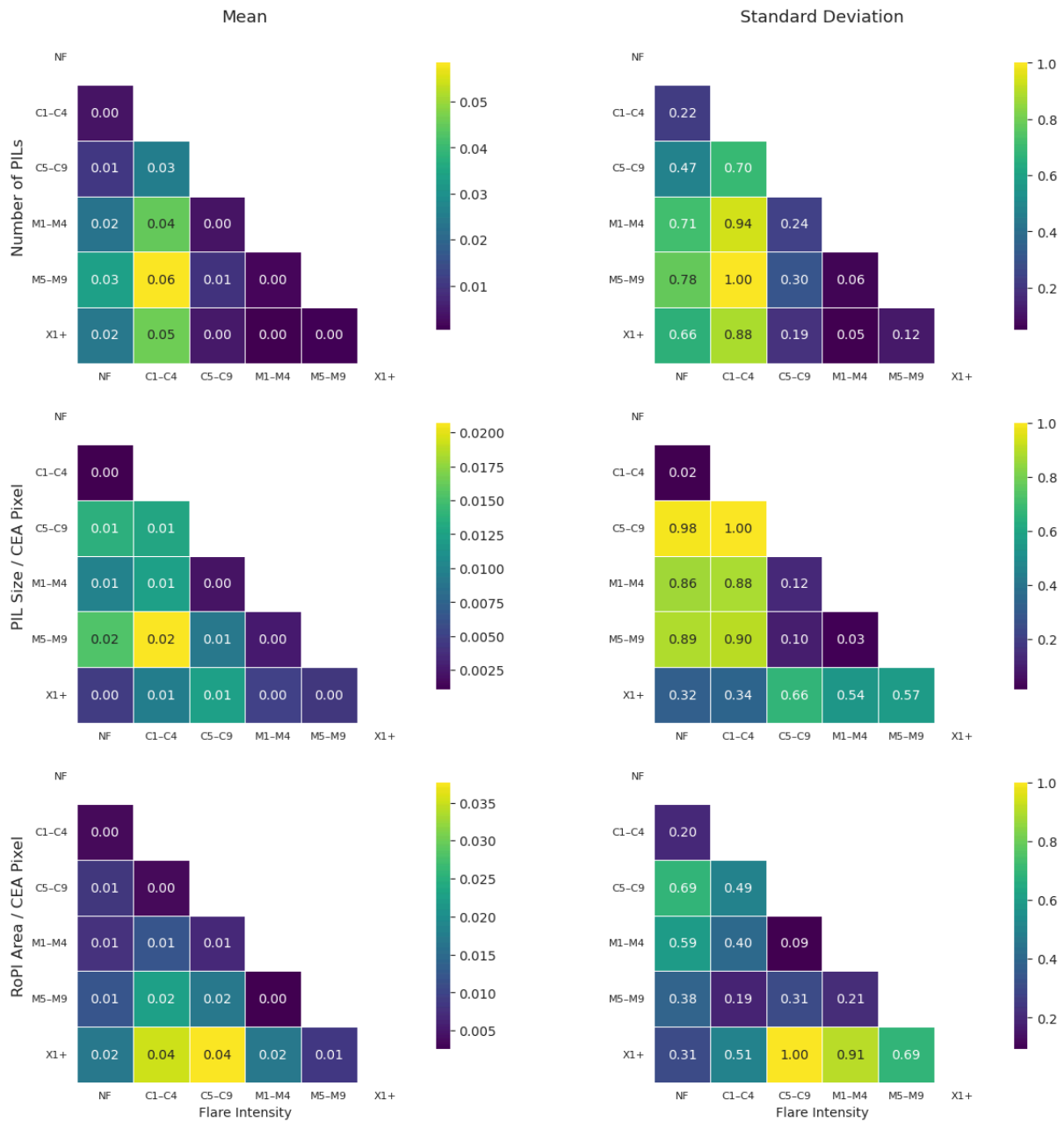

**Fig A4.** Pairwise Bhattacharyya distances between flare class categories for MPIL count (top row), MPIL size (middle row), and RoPI area (bottom row) in the ±250 G threshold tier. Left column: distances based on mean differences; right column: distances based on standard deviation differences. Distances are lower than for ±50 G and ±100 G, reflecting reduced feature variability and separability at this higher threshold.

### A.3 ±500 G Threshold

**Distribution of MPIL Features by Flare Class: Fig. A5** shows the distributions of MPIL size, MPIL count, and RoPI area for the ±500 G threshold.
 At this strict cutoff, only the most intense core segments of polarity inversion lines are detected.

   MPIL Size values are tightly concentrated between ~$10^2$ and $3\times10^2$ pixels, with minimal variation across flare classes.
   MPIL Count is almost always 1, regardless of flare class, eliminating its ability to discriminate between categories.
   RoPI Area retains some variability, with slightly higher medians for certain C- and M-class events, but overall trends with flare class are weak and irregular.
   Compared to lower thresholds, this high threshold discards most mid-field structures, significantly reducing the observable morphological differences between flaring and non-flaring regions.

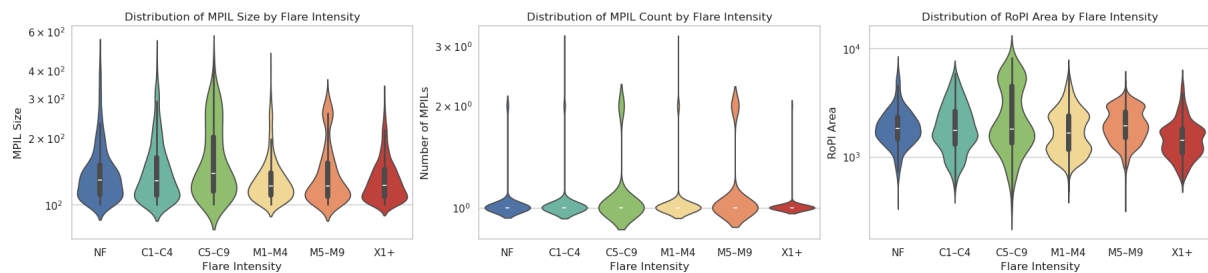

**Fig A5.** Violin plots of MPIL size (left), MPIL count (center), and RoPI area (right) across flare class categories for the ±500 G threshold. Distributions are shown on a logarithmic scale; white dots represent medians, thick bars indicate interquartile ranges, and internal ticks show local density. At this threshold, features are dominated by the strongest core segments, with minimal variation across classes.

**Bhattacharyya Distance Analysis Fig. A6** presents the Bhattacharyya distance matrices for the ±500 G threshold. Distances are small for MPIL Size and MPIL Count across all class pairs, confirming their lack of discriminative power at this threshold. RoPI Area produces slightly higher distances, particularly between NF and the strongest flare classes, but still much lower than at ±50 G or ±100 G.
   Standard deviation–based distances show that variability differences across flare classes are muted, reflecting the uniformity introduced by selecting only the most extreme field regions.
   These results suggest that overly strict thresholds, such as ±500 G, can severely limit the utility of MPIL features for flare classification.

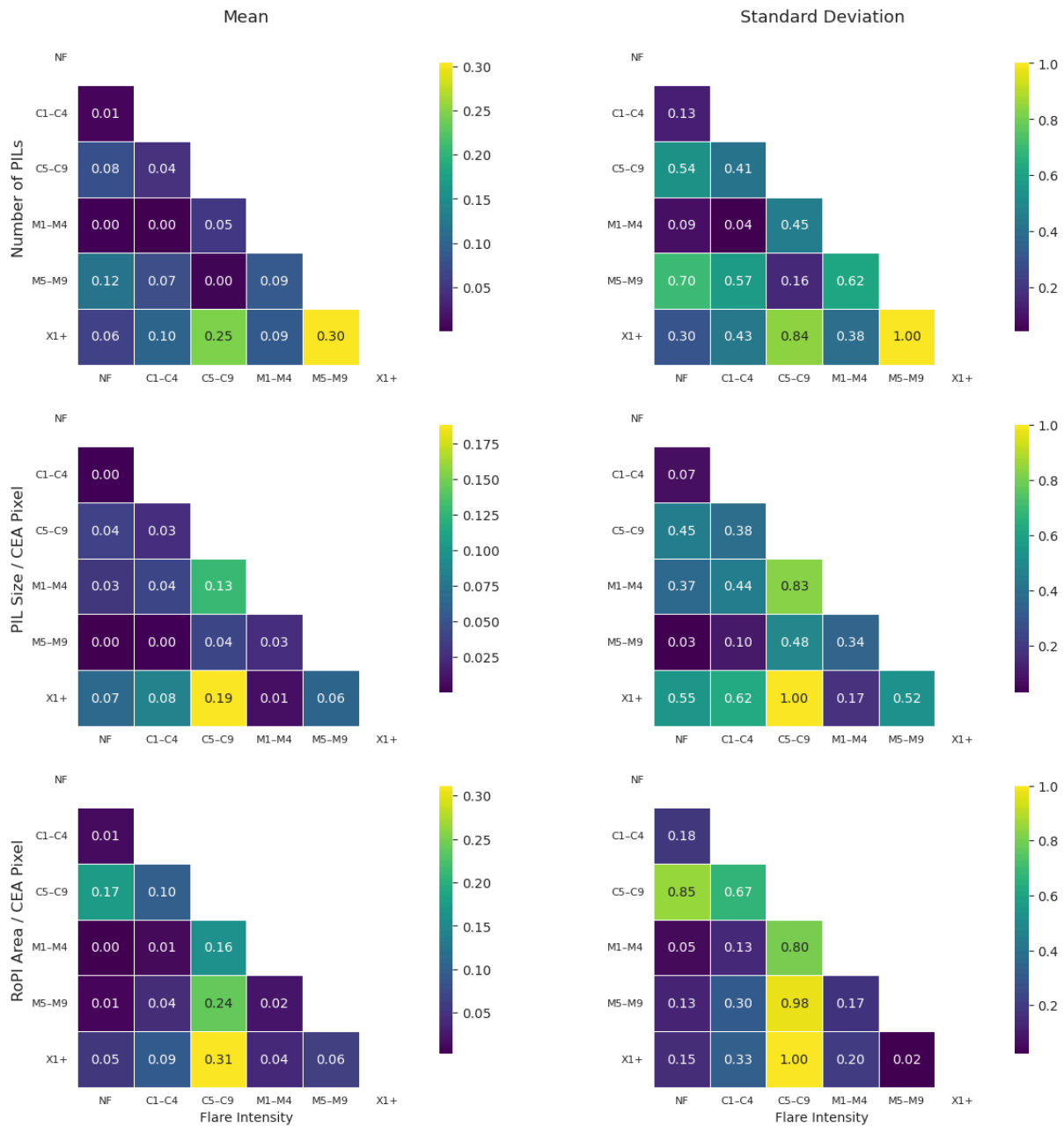

**Fig A6.** Pairwise Bhattacharyya distances between flare class categories for MPIL count (top row), MPIL size (middle row), and RoPI area (bottom row) in the ±500 G threshold tier. Left column: distances based on mean differences; right column: distances based on standard deviation differences. Distances are small across most class pairs, indicating limited separability at this high threshold.

Across all tested thresholds, the ±100 G baseline strikes the best balance between capturing sufficient MPIL structure for discriminative power and avoiding excessive noise from weak-field regions. The ±50 G threshold yields larger feature magnitudes and still shows clear trends with flare class. Still, the inclusion of weak-field segments increases within-class variance and reduces separability between adjacent categories. Increasing the cutoff to ±250 G and ±500 G progressively compresses feature ranges, suppresses variability, and diminishes class separation, especially for MPIL Count, which becomes almost uniform at the highest threshold. These results highlight that intermediate thresholds (around ±100 G) preserve both strong and mid-field MPIL structure, making them more informative for flare-related morphological analysis.